\documentclass{emulateapj}
\usepackage{float, amsmath,verbatim,epsf}

\shorttitle{Particle acceleration in relativistic collisionless shocks} 
\shortauthors{Spitkovsky}

\begin{document}

\title{Particle acceleration in relativistic collisionless shocks: Fermi process at last?} 
\author{Anatoly Spitkovsky\altaffilmark{1}}
\altaffiltext{1} {Department of Astrophysical Sciences, Peyton Hall,
Princeton University, Princeton, NJ 08544: anatoly@astro.princeton.edu}

\subjectheadings{acceleration of particles -- collisionless shocks -- gamma rays: bursts }
\begin{abstract}

We present evidence that relativistic shocks propagating in unmagnetized plasmas can self-consistently accelerate particles. We use long-term two-dimensional particle-in-cell simulations to study the well-developed shock structure in unmagnetized pair plasma. The particle spectrum downstream of such a shock consists of two components: a relativistic Maxwellian, with characteristic temperature set by the upstream kinetic energy of the flow, and a high-energy tail, extending to energies $>100$ times that of the thermal peak. This tail is best fitted as a power law in energy with index $-2.4\pm0.1$, modified by an exponential cutoff. The cutoff moves to higher energies with time of the simulation, leaving a larger power law range. 
The number of particles in the tail is $\sim1\%$ of the downstream population, and they carry $\sim 10\%$ of the kinetic energy in the downstream.  
Upon investigation of the trajectories of particles in the tail, we find that the energy gains occur as particles bounce between the upstream and downstream regions in the magnetic fields generated by the Weibel instability. We compare this mechanism to the first order Fermi acceleration, and set a lower limit on the efficiency of shock acceleration process. 
\end{abstract}
\maketitle
\section{Introduction}

Acceleration of particles in collisionless shocks is at the heart of 
most models of nonthermal phenomena in the Universe. Observations of synchrotron emission from astrophysical sources suggest that collisionless shocks in pulsar wind nebulae (PWNe), 
jets from active galactic nuclei (AGNs), gamma-ray bursts (GRBs) and supernova remnants 
(SNRs) can convert a significant fraction of the flow energy into relativistic particles with power law nonthermal spectra. This is usually attributed to the first-order Fermi mechanism -- a process in which particles scatter between the upstream and downstream regions of shocks to gain energy (see Blandford \& Eichler 1987 for review). Yet, despite its significance in astrophysics, the Fermi mechanism has not been demonstrated to work self-consistently from first principles, and its efficiency and conditions for operation are not well understood. Most progress in studying shock acceleration has been made using Monte Carlo test particle simulations (e.g., Ostrowski \& Bednarz 1998, Ellison \& Double 2004) and semi-analytic kinetic theory methods (e.g., Kirk et al. 2000, Achterberg et al. 2001, Keshet \& Waxman 2005). Both methods study particle acceleration with certain assumptions about the scattering processes near shocks, which in turn depend on the nature of magnetic turbulence in the flow. Whether realistic shock turbulence leads to particle acceleration is currently unknown. This is especially true in the case of relativistic shocks, where observational constraints on the turbulence properties are lacking. The injection efficiency, or the fraction of particles in the flow that are nonthermally accelerated, may depend on the details of the shock transition region and is not well constrained.  
Resolving these issues requires a self-consistent calculation that can simultaneously capture the physics of the shock-generated turbulence and the particle acceleration processes. In this Letter, we demonstrate via ab-initio particle-in-cell (PIC) simulations that relativistic collisionless shocks propagating in initially unmagnetized electron-positron pair plasmas naturally produce accelerated particles as part of the shock evolution. These particles form a power law tail, and we argue that the acceleration process is very similar to Fermi acceleration. 
We are thus able to measure the efficiency of shock acceleration without any assumptions. 
The particular case of unmagnetized relativistic shocks that we consider here is relevant to the prompt and afterglow emission from GRBs (e.g., Waxman 2006). This is also one of the cleanest tests of the shock acceleration theory because in an initially unmagnetized flow all magnetic turbulence has to be self-generated. 
In \S\ref{secsetup} we describe the simulations and shock structure and then discuss the particle acceleration mechanism in \S\ref{secaccel}. 
 
\section{Collisionless shock structure}\label{secsetup}

Collisionless shocks in unmagnetized relativistic flows are mediated by the Weibel instability (Weibel 1959; Medvedev \& Loeb 1999; Gruzinov \& Waxman 1999), which converts the free energy of anisotropic streaming of interpenetrating flows into small scale (skindepth) magnetic fields. The fields grow to subequipartition levels and deflect and randomize the bulk flow, creating the shock compression (Kato 2005, Milosavljevic et al. 2006). PIC simulations of colliding relativistic shells have confirmed the general picture of the initial stages of the instability (Kazimura et al. 1998; Nishikawa et al. 2003, 2005; Silva et al. 2003; Frederiksen et al. 2004; Hededal et al. 2004), and have been evolved through the shock formation stage in both  2D (Gruzinov 2001; Kato 2007; Spitkovsky 2008, hereafter S08; Chang et al. 2008, herefter CSA08) and 3D (Spitkovsky 2005, hereafter S05). A persistent feature of all unmagnetized shock simulations had been the downstream particle spectrum that is well approximated by a relativistic Maxwellian (S05; Kato 2007). Reports of nonthermal spectral components in short 3D simulations (Nishikawa et al. 2003, Hededal et al. 2004) can probably be attributed to the incomplete flow thermalization before a shock fully forms. Long simulations, as presented in this Letter, show that additional nonthermal spectral components develop over time. 

We use the electromagnetic PIC code {\it TRISTAN-MP} (Buneman 1993; S05) to simulate a relativistic shock propagating through an unmagnetized $e^\pm$ plasma. The shock is triggered by reflecting an incoming cold ``upstream" flow that propagates  with Lorentz factor $\gamma_0=15$ in the $-x$ direction from a conducting wall at $x=0$. The simulation is performed in the ``wall" or ``downstream" frame (S05; S08). To maximize  feasible simulation size we use the 2D version of {\it TRISTAN-MP} with flow direction lying in the simulation plane. Although we track all three components of velocity, in the case of an unmagnetized 2D shock only the in-plane velocities and currents are excited. The only magnetic field component is then the out-of-plane $B_z$ (CSA08). 
\begin{figure} 
\includegraphics[scale=.54]{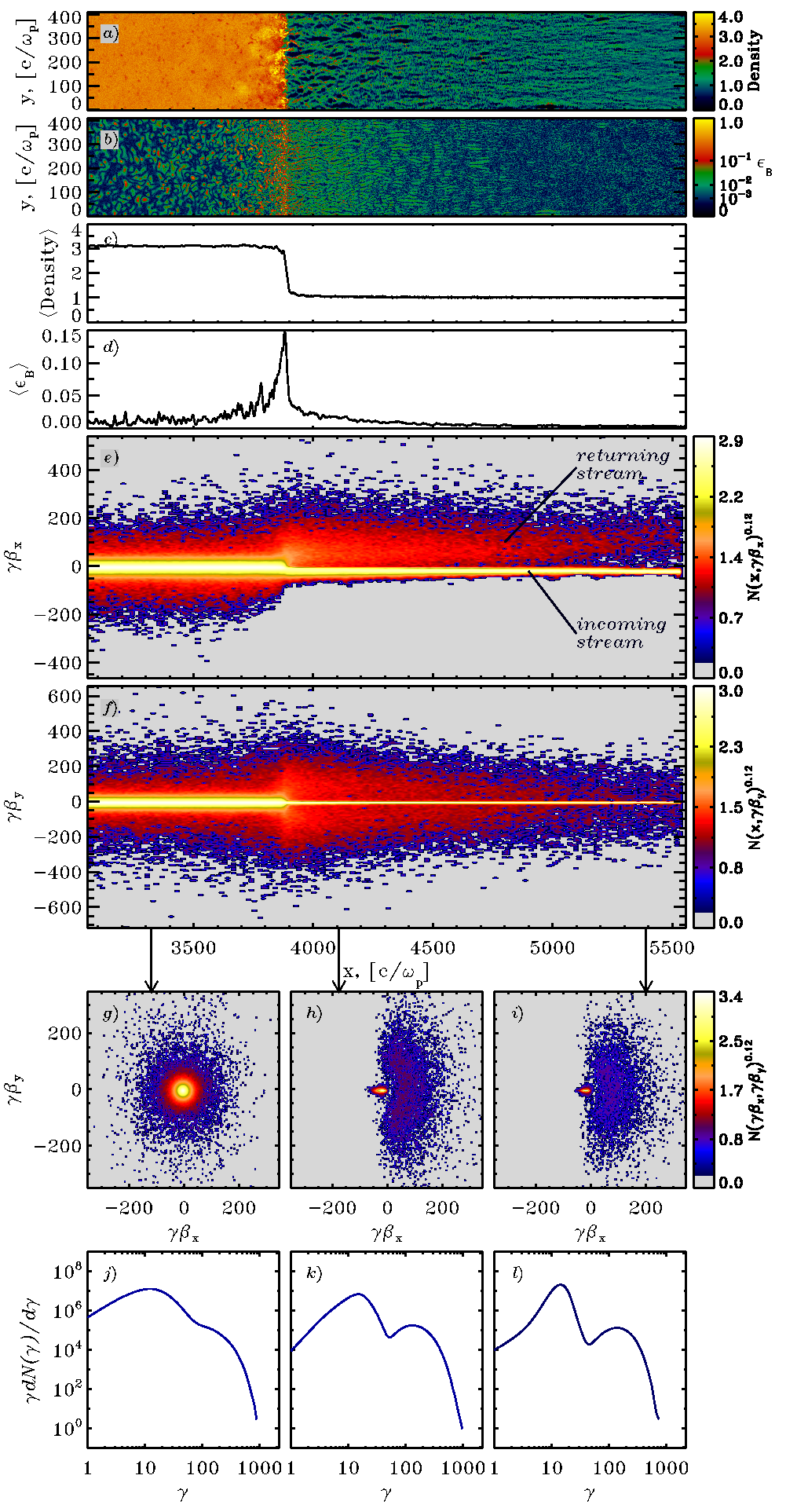}
\vskip -.05in
\caption{Internal structure of a relativistic pair shock at $\omega_p t=8400$:
a) Density in the simulation plane (scaled by the upstream density); b) Magnetic energy density $\epsilon_B\equiv B^2/4\pi \gamma_0 n_1 m c^2$; c) Transversely averaged plasma density; d) Transversely averaged  $\epsilon_B$; e) Longitudinal phase space for electrons; f) Transverse phase space for electrons; g-i) Momentum space $\gamma \beta_x - \gamma \beta_y$ at three slices through the flow (locations marked by arrows); j-l) Electron spectra in corresponding slices.
}
\label{Fig1}
\end{figure}

Our numerical parameters are as follows: the skindepth based on the upstream density $n_1$ is $c/\omega_p=(4 \pi e^2 n_1/\gamma_0 m_e c^2)^{-1/2}=10$ cells, the transverse box size $l_y=400 c/\omega_p$, and timestep $\omega_p\Delta t  = 1/22$, with 8 particles/cell/species in the incoming flow. We have also tried larger box sizes ($l_y\leq1024 c/\omega_p$) and more particles/cell/species (up to 32) with no significant changes to the results. The simulations are evolved for times $\gtrsim 10^4 \omega_p^{-1}$. The final box length is $l_x \gtrsim 10^4 c/\omega_p$.  

In Fig. \ref{Fig1} we show the internal structure of the shock 
at time $\omega_p t=8400$. Density (Fig. \ref{Fig1}a) and magnetic field energy (Fig. \ref{Fig1}b) show the Weibel filamentation in the upstream region, with characteristic transverse filament scale $\sim 20 c/\omega_p$. The filaments are advected with the upstream flow into the shock and undergo merger. The flow rapidly decelerates over $\sim 50 c/\omega_p$, and the density increases to $n_2/n_1=3.13$ (Fig. \ref{Fig1}c), as given by the jump conditions for a 2D relativistic gas measured in the downstream frame (CSA08; S08). The shock moves to the right at $0.47c$ (shock Lorentz factor with respect to the upstream is $\Gamma_{shock}=\sqrt{2} \gamma_0=21.2$). Transversely averaged magnetic energy reaches a peak at $\sim 15 \%$ of equipartition and then decays (Fig. {\ref{Fig1}d; CSA08). In Fig. \ref{Fig1}e, we show the longitudinal momentum phase space $x-\gamma \beta_x$ for electrons, where the phase space density of particles is shown as a 2D histogram (positron phase space is nearly identical). The incoming stream appears at a negative 4-velocity $\gamma \beta_x \approx -15$, and the flow stops and thermalizes after the shock. In the upstream, there is a clear population of returning particles with positive 4-velocity (Milosavljevic et al. 2006; Kato 2007; S08). 
Such hot, low density particles moving in front of the shock drive filamentation in the upstream and are essential for maintaining the shock as a self-propagating structure far from the wall. 
Phase space snapshots $\gamma \beta_x - \gamma \beta_y$ at three cross sections through the flow are shown in Fig. \ref{Fig1}g-i. The downstream distribution displays near isotropy, while in the upstream the returning particles are anisotropic, with an excess of high-energy particles that move at large angles to shock normal. 

In Fig. {\ref{Fig1}j-l} we show particle spectra in the downstream and the upstream regions, measured in the reference frame of the downstream medium. In the upstream, the returning particles appear as a high-energy peak in addition to the incoming flow. In the downstream, the spectrum consists of the main peak, centered around $\gamma\approx12$, and a high energy ``tail" extending to $\gamma\approx1000$. This tail develops in all our runs at times $\omega_p t \gtrsim 10^3$. In Fig. \ref{Fig2} we show a fit to the spectrum from a slice centered at $500 c/\omega_p$ downstream from the shock at time $\omega_p t=10^4$. The fit consists of a relativistic 2D Maxwellian plus a power law with an exponential cutoff of the form $f(\gamma)=C_1 \gamma \exp(-\gamma/ \Delta \gamma_1) + C_2 \gamma^{-\alpha} \min[1,\exp(-(\gamma-\gamma_{cut})/\Delta \gamma_{cut})]$, with normalizations $C_1$ and $C_2$, such that $C_2=0$ for $\gamma<\gamma_{min}$. In the fit, the Maxwellian spread is $\Delta \gamma_1=6$ (close to but smaller than $\Delta \gamma=(\gamma_0-1)/2=7$ expected from the full thermalization of the upstream flow). The power law begins at $\gamma_{min}=40$ with index $\alpha=2.5$, and the high energy cutoff starts at $\gamma_{cut}=300$ with a spread $\Delta \gamma_{cut}=100$. For comparison, the red line in Fig. \ref{Fig2}a shows a fit with two Maxwellians, demonstrating a clear deficit at intermediate energies, while fitting well the low and high energy ends. The power law stretch develops over time, as can be seen in Fig. \ref{Fig2}b, where a sequence of spectra measured at a fixed distance behind the shock is shown at different times. The high energy cutoff and maximum energy in the tail grow with time and the power law range expands. The power law index varies between $2.3-2.5$ (smaller values are obtained if the fit is not required to match the transition region between the Maxwellian and the power law). At $\omega_p t=10^4$, the tail at $\gamma > 75$ contains $\sim 1\%$ of particles and $\sim 10\%$ of energy in the downstream. 
\section{Acceleration mechanism}\label{secaccel}
We studied the mechanism that populates the suprathermal tail by tracing the orbits of particles that gain the most energy. 
The main acceleration happens near the shock, as seen from the excess of particles with large 4-velocity near the shock in Fig. \ref{Fig1}e. 
The space-time trajectory $x(t)-x_{shock}(t)$ and the acceleration history $\gamma(t)$ for four representative particles are shown in Fig. \ref{Fig3}. The vast majority of particles in the flow go through the shock only once and never return to the upstream again (orange line in Fig. \ref{Fig3}). Some, however, can cross the shock several times and gain energy. 
After acceleration near the shock, these particles escape into the upstream or the downstream, populating the suprathermal tails (red, green and blue particles). 
The particles that gain the most energy (red and blue lines) undergo several reflections between the downstream (or the shock layer) and the upstream, with the largest energy gains coming from reflections in the upstream region (Fig. \ref{Fig3}). Upon each reflection these particles gain energy $\Delta E \sim E$, as expected in relativistic shocks. In Fig. \ref{Fig3} we overplot the transversely averaged magnetic energy as line plots stacked in time. Note, that all quantities are still measured in the downstream frame and are only shifted in space so that the shock appears stationary. Magnetic fluctuations associated with the upstream filaments carry motional electric field ($E_y$) as they are advected towards the shock. Particles moving against the flow in these fields scatter with a net energy gain (in contrast, deflections in the downstream region result in no energy gain as seen in the downstream frame). 
The shock width of $\sim 50 c/\omega_p$ is roughly equal to several Larmor radii of the thermal particles in the self-generated field. The accelerated particles, however, have Larmor radii that significantly exceed the width of the shock. A high energy particle moving along the shock normal will not reflect at the shock, and will leave through the downstream, where the fields are weaker. Instead, the particles that get accelerated to the highest energies move almost parallel to the shock surface, across the magnetic filaments. For them, the alternating magnetic polarity of the filaments represents magnetic fluctuations on scales smaller than the Larmor radius. The deflections in the upstream toward the downstream are thus grazing incidence collisions with moving magnetic islands. The deflections towards the upstream happen within several hundred skindepths behind the shock, where the magnetic field is strongest during the simulation. 

\begin{figure} 


\centerline{\includegraphics[scale=.54]{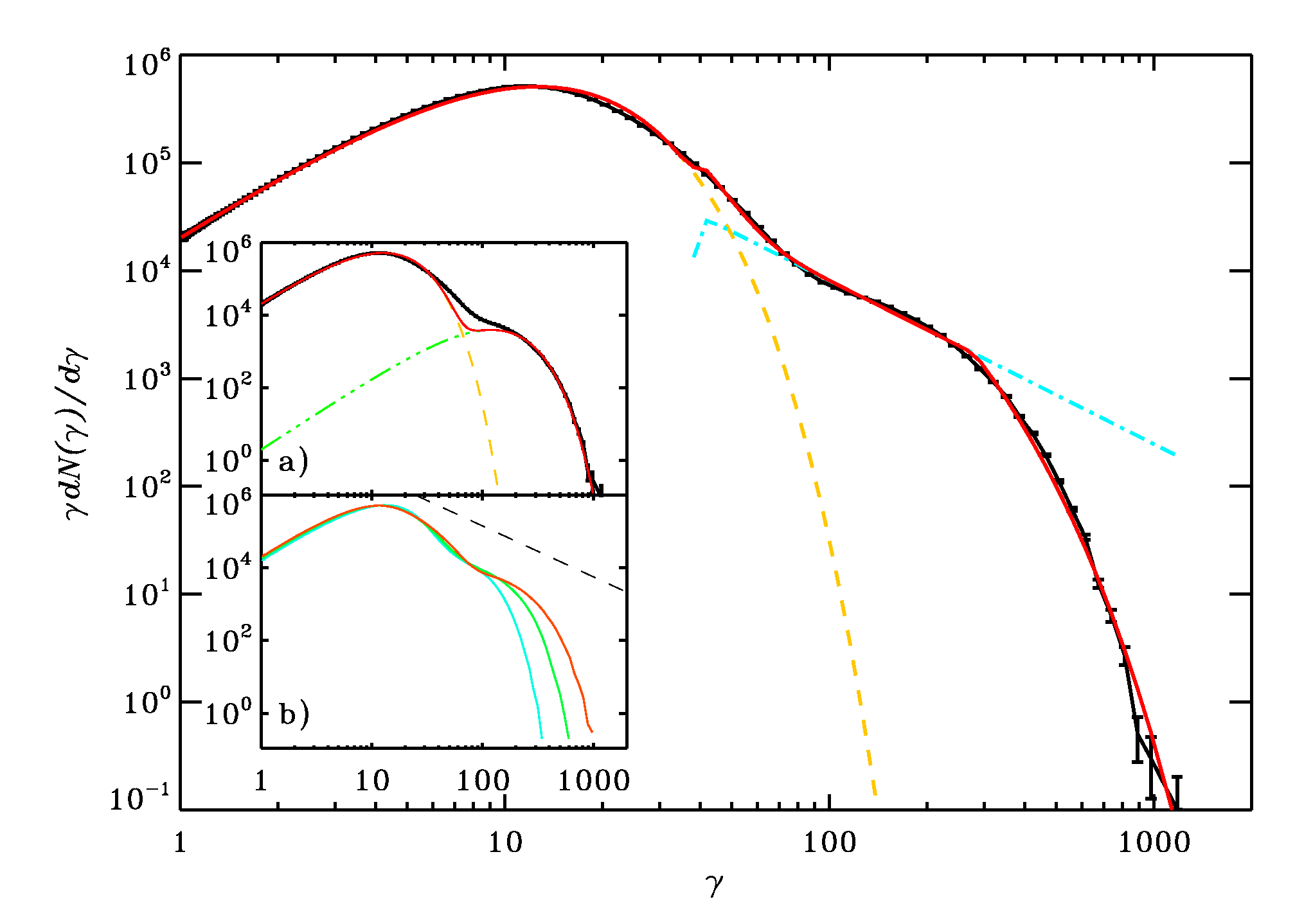}}
\vskip -.1in
\caption{Main panel: Particle spectrum in a $100 c/\omega_p$-wide slice at $500 c/\omega_p$ downstream from the shock at time $\omega_p t = 10^4$ (black line with error bars). Red line: a fit with a 2D Maxwellian (yellow dashed) plus a power law (blue dash-dot) with high-energy exponential cutoff. Subpanels:  a) fit with a sum of high and low temperature Maxwellians (red line), showing a deficit at intermediate energies; b) evolution of particle spectrum in a downstream slice with time: $1600\omega_p^{-1}$ (blue),  $3800\omega_p^{-1}$ (green), $10^4\omega_p^{-1}$(red). Black dashed line shows $\gamma^{-2.4}$ power law. }
\label{Fig2}
\end{figure}

\begin{figure} 
\vskip -.17in
\centerline{\includegraphics[scale=.54]{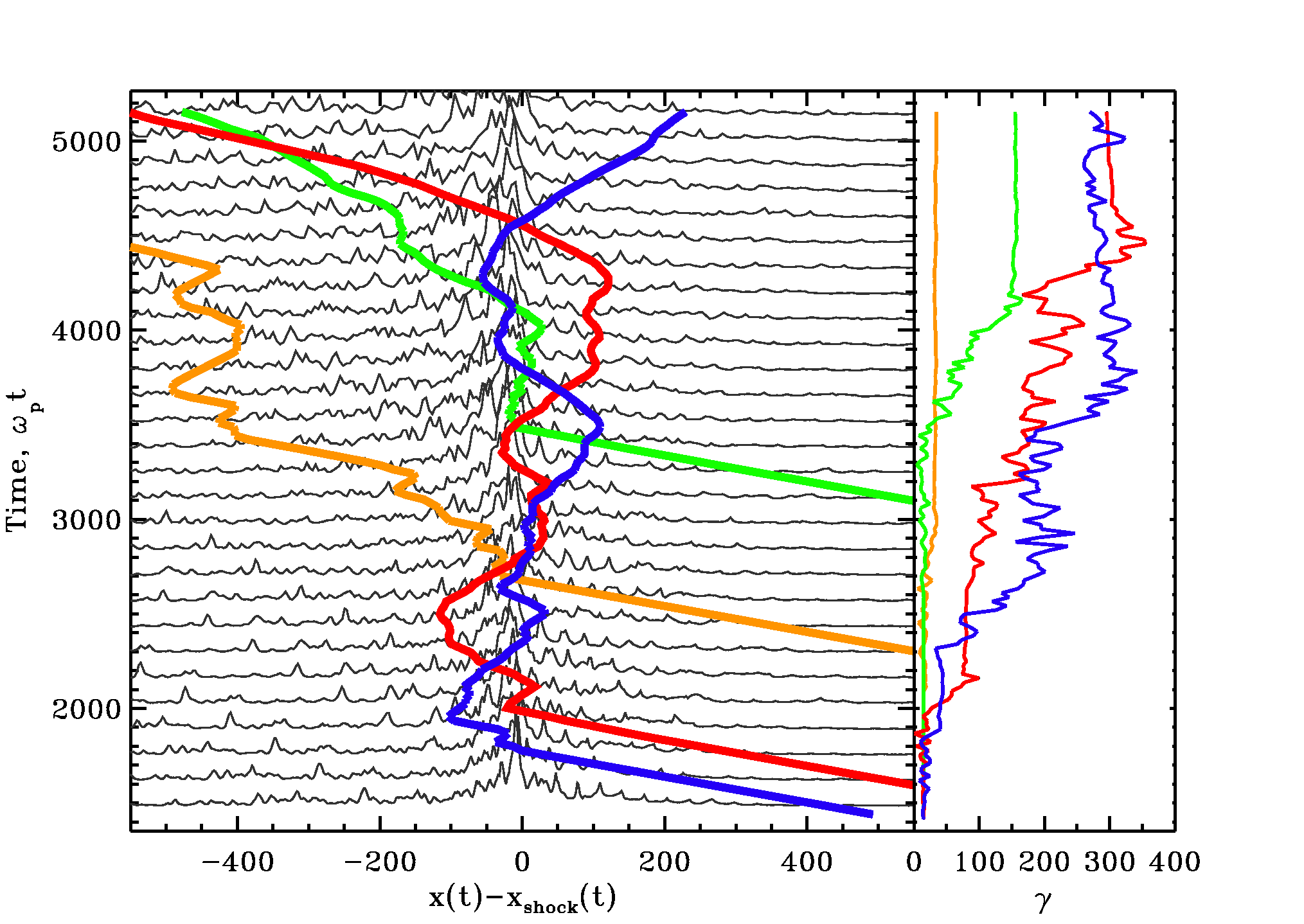}}
\vskip -.05in
\caption{Left panel: horizontal position as a function of time for four representative particles (color lines) overplotted on transversely averaged profiles of magnetic energy (gray lines). Right panel: particle energies (shown with corresponding color lines) as a function of time. All horizontal positions are shifted by $x_{shock}(t)$ to align them with the shock location. All quantities are measured in the downstream frame.}
\label{Fig3}
\end{figure}

\section{Discussion}

We have shown that relativistic collisionless shocks can self-consistently accelerate nonthermal particles. The magnetic fields that are created as part of the Weibel turbulence near shocks are sufficient to inject particles from the thermal pool into the shock acceleration process, and our findings do not require any initial assumptions about the turbulence spectrum. The acceleration we observe involves repeated crossings of the shock by a small fraction of particles. In this sense, it is tempting to associate this acceleration process with the first order Fermi acceleration in shocks.  The absence of a coherent background field in our problem rules out shock-surfing and shock-drift acceleration. Deviations in particle trajectory are due to interactions with many magnetic filaments of alternating magnetic polarity and is, therefore, diffusive in nature. 
The exponential cutoff at high energy in the downstream spectrum can be attributed to the finite acceleration time in the simulation and the fact that efficient particle scattering occurs in a layer of finite width (Bykov \& Uvarov 1999). We have not reached a steady state, however. Both the maximum particle energy and the extent of the particle precursor continue to grow linearly with time, and the energy in the tail grows logarithmically. The region where particles scatter also increases. We 
conclude that the simulations show the beginnings of the Fermi acceleration process, and larger simulations may display further evolution of the power law tail. 
 However, even now we can place a lower limit on the efficiency of shock acceleration: at least $\sim1\%$ of particles are injected into the nonthermal tail. 

The amount of energy in the tail at the end of our simulation is significant enough ($\sim 10\%$) to affect the jump conditions at the shock. The Maxwellian part of the downstream spectrum appears at a lower temperature than would be expected from simple thermalization of the upstream kinetic energy. Yet, the mean energy per particle in the downstream region is still equal to the kinetic energy per particle far upstream. The difference is due to the contribution from the suprathermal tail to the energy balance. In fact, in addition to the ``cooling" of the Maxwellian component, we observe that the fit by a Maxwellian plus power law becomes worse with time, mainly because the low-energy peak becomes more distorted, with more higher-energy particles than predicted by a pure Maxwellian. This transition region may require a different spectral component. Extrapolating this process, we expect that the eventual fraction of power law particles may exceed few percent, and the thermal peak may get significantly eroded. This process may explain the absence of a clear signature of postshock thermal emission in PWN and GRB shocks. 

The acceleration process described in this Letter is very robust. The development of the tail is not sensitive to the transverse size of the box (as long as at least several Weibel filaments can fit into the box), or to the varying $\gamma$ of the upstream flow (we tried $\gamma_0=2$ to $100$). We have not studied nonrelativistic shocks, because the electromagnetic Weibel instability may be less efficient in that case, and the shock structure may change. This is the subject of future work. In the electron-ion relativistic shocks, however, the beginnings of the tail formation are already clear in the unmagnetized simulations with mass ratio $m_i/m_e=100$, $\gamma_0=15$ in S08. It is commonly expected that electrons have more difficulty being injected into the Fermi process, because their Larmor radii have to be comparable to the shock thickness that is set by ion dynamics. The efficient energy transfer between ions and electrons found in S08 eliminates this problem, and the relativistic electron-ion shocks thus behave very similarly to electron-positron shocks. Therefore, given sufficient computational resources we expect to see nonthermal tails in both electrons and protons, accelerated by the Fermi process. Although our simulations are two-dimensional, we expect the same process to work in three-dimensional shocks. We were able to run 3D pair shock simulations for $2000\omega_p^{-1}$, and have seen the beginnings of nonthermal tail formation very similar to the 2D case (Spitkovsky \& Arons, in preparation).

Our results underscore the importance of long-term PIC simulations for characterizing both the collisionless shock structure and the nonlinear processes that naturally appear once a self-consistent shock structure is in place. Here we addressed the acceleration of particles from the thermal pool. The effect of such accelerated particles on the evolution of magnetic fields in the shock will be presented elsewhere (Keshet et al. 2008). The self-consistent picture that emerges from PIC shock simulations suggests that the astrophysical scenario of shocks depositing $\sim10\%$ of energy into nonthermal particles is viable on the microphysical level. In application to GRBs our results suggest that prompt and afterglow shocks can have nonthermal electron energy fraction of $\epsilon_e\approx 0.1$. For PWNe, the mechanism presented here can work in shocks near the rotational equator of a pulsar, where, due to the reconnection of oppositely directed toroidal fields, the local value of magnetization can be small enough to allow Weibel-driven shocks. 
\vskip 0.06in
We thank J. Arons, R. Blandford, U. Keshet, \& E. Waxman for help and suggestions. We acknowledge the use of {\it TIGRESS} computing center at Princeton and the support from Alfred P. Sloan Foundation fellowship. 

\references

\noindent
Achterberg, A., Gallant, Y.~A., Kirk, J.~G., Guthmann, A.~W. 2001, \mnras, 328, 393 

\noindent
Blandford, R. D. \& Eichler D., 1987, Phys. Rep, 154, 1

\noindent
Buneman, O. 1993 in ``Computer Space Plasma Physics'', Terra Scientific, Tokyo, 67

\noindent
Bykov A. M. \& Uvarov, Yu. A. 1999, {\it JETP}, 88, 465

\noindent
Chang, P., Spitkovsky, A., \& Arons, J. 2008, \apj, 674, 378 (CSA08)
\noindent
Ellison, D. C. \& Double, G. P. 2004, Astropart. Phys., 22, 323

\noindent
Frederiksen, J. T., Hededal, C. B., Haugb{\o}lle, T., \& Nordlund, \AA. 2004, \apjl, 608, L13

\noindent
Gruzinov, A. \& Waxman E. 1999, \apj, 511, 852

\noindent
Gruzinov, A. 2001, astro-ph/0111321

\noindent 
Hededal, C.~B., Haugb{\o}lle, T., Frederiksen, J.~T., \& Nordlund, {\AA}. 2004, \apj, 617, L107

\noindent
Kato, T.~N. 2005, {\it Phys. Plasmas}, 12, 80705

\noindent
Kato, T.~N. 2007, \apj, 668, 974

\noindent
Kazimura Y., Sakai, J. I., Neubert, T., Bulanov, S. V. 1998, {\apjl}, 498, L183

\noindent 
Keshet, U. \& Waxman E. 2005, {\it Phys. Rev. Lett.}, 94, 111102

\noindent
Keshet, U., Katz, B., Spitkovsky, A. \& Waxman, E. 2008, submitted, arXiv:0802.3217

\noindent
Kirk, J.~G., Guthmann, A.~W., Gallant, Y.~A., Achterberg, A. 2000, \apj, 542, 235
	
\noindent
Medvedev, M.~V. \& Loeb, A. 1999, \apj, 526, 697

\noindent
Milosavljevic, M., Nakar, E., \& Spitkovsky, A. 2006, \apj, 637, 765

\noindent
Nishikawa, K.-I., Hardee, P., Richardson, G., Preece, R., Sol, H., \&
Fishman, G. J. 2003, \apj, 595, 555

\noindent
---------- 2005, \apj, 622, 927

\noindent
Ostrowski M. \& Bednarz J. 1998, {\it Phys. Rev. Lett.}, 80, 18

\noindent
Silva, L. O., Fonseca, R. A., Tonge, J. W., Dawson, J. M., Mori,
W. B., \& Medvedev, M. V. 2003, \apj, 596, 121

\noindent
Spitkovsky, A. 2005, AIP Conf. Proc, 801, Astrophysical Sources of High Energy Particles and Radiation, ed. T. Bulik, B. Rudak, \& G. Madejski (New York: AIP) 345 (S05); astro-ph/0603211

\noindent
Spitkovsky, A. 2008, \apjl, 673, L39 (S08)

\noindent
Waxman, E. 2006, Plasma Phys. Control. Fusion 48, B137

\noindent
Weibel, E. S. 1959, {\it Phys. Rev. Lett}, 2, 83

\end{document}